\documentclass[conference]{IEEEtran}
\makeatletter
\def\ps@headings{%
\def\@oddhead{\mbox{}\scriptsize\rightmark \hfil \thepage}%
\def\@evenhead{\scriptsize\thepage \hfil \leftmark\mbox{}}%
\def\@oddfoot{}%
\def\@evenfoot{}}
\makeatother
\usepackage{epsfig}
\usepackage{graphicx}
\usepackage{amsmath}
\usepackage{algorithmic}
\usepackage{multirow}

\begin{document}

\title{ A Novel Association Policy for Web Browsing in a Multirate WLAN }

\author{Pradeepa BK and Joy Kuri \\ Centre for Electronics Design and Technology, \\
Indian Institute of Science, Bangalore. India. \\
{bpradeep,kuri}@cedt.iisc.ernet.in
}
\maketitle
\begin{abstract}
We obtain an association policy for STAs in an IEEE 802.11 WLAN
by taking into account explicitly two aspects of practical importance:
(a) TCP-controlled short file downloads interspersed with read times
(motivated by web browsing), and (b) different STAs associated with
an AP at possibly different rates (depending on distance from the AP).

Our approach is based on two steps. First, we consider an analytical
model to obtain the aggregate AP throughput for long TCP-controlled
file downloads when STAs are associated at $k$ different rates $r_1$,
$r_2$, $\ldots$, $r_k$; this extends earlier work in the literature.
Second, we present a 2-node closed queueing network model to approximate
the expected average-sized file download time for a user who shares the
AP with other users associated at a multiplicity of rates.
These analytical results motivate the proposed association policy,
called the Estimated Delay based Association (EDA) policy:
Associate with the AP at which the expected file download
time is the least.

Simulations indicate that for a web-browsing type traffic scenario,
EDA outperforms other policies that have been proposed earlier; 
the extent of improvement ranges from 12.8\% to 46.4\%
 for a 9-AP network.
To the best of our knowledge, this is the first work that proposes
an association policy tailored specifically for web browsing.
Apart from this, our analytical results could be of independent
interest. 
\end{abstract}

\IEEEpeerreviewmaketitle

\section{Introduction}\label{sec:Introduction}
%I wish you the best of success.

\begin{figure}
\centering
\includegraphics[scale=0.7]{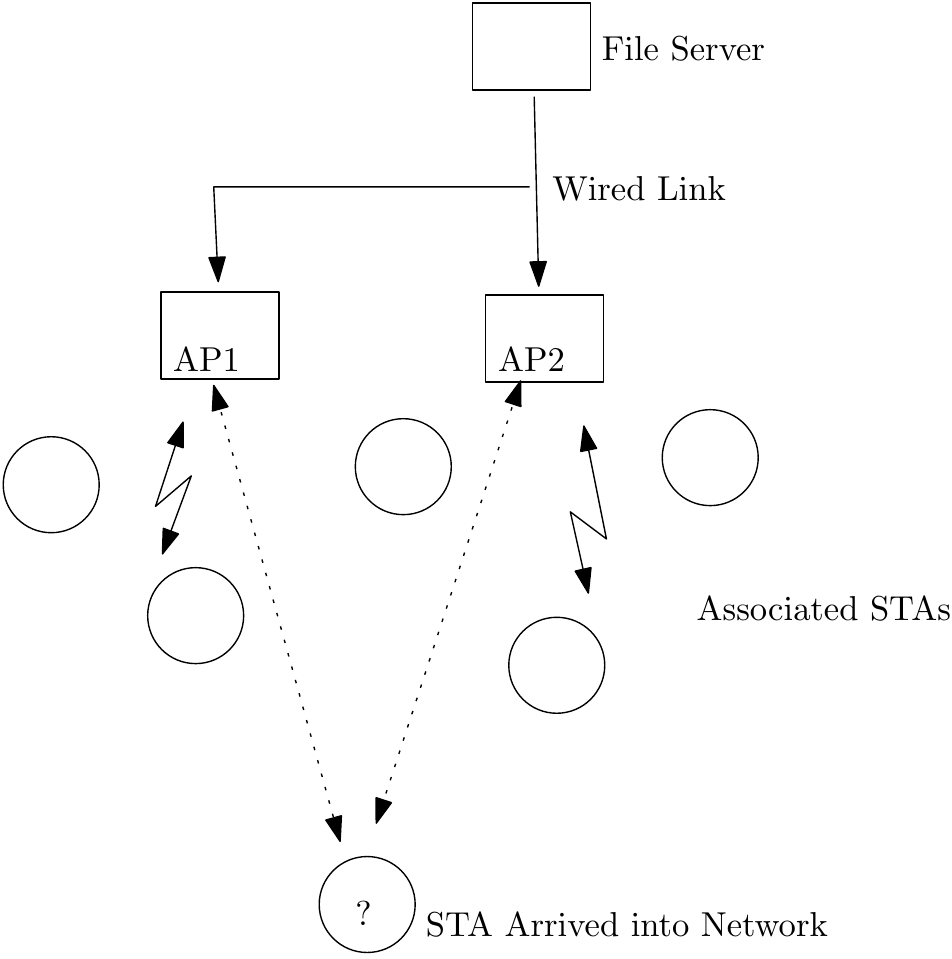} 
\caption{Associated STAs are downloading the files from a server through 
APs and a new STA arrived to region, where it can potentially get associated 
with any one of the APs}
\label{fig:arrived_STA}
\end{figure}

IEEE 802.11a/b/g/n based Wireless LAN are used for providing Internet access in
many places. In most deployed network a single station (STA) can see multiple access
points (APs) as shown in Figure \ref{fig:arrived_STA}. In this situation, STAs associate with a
particular AP. Usually, association of STA with AP is based on SNR. This simple
scheme ignores the fact that some APs are loaded heavily while some 
are underutilized. Hence, load imbalance and lower throughput are some
 common problems. Prior and contemporary works address these problems by
assuming a \emph{saturated} traffic model. They  assume that  at any
point of time, every STA is hungry to send packets across and is always 
contending for the channel. However, according to the statistics given in
 \cite{astn_model:Businesswire}, 46\% 
of Internet traffic is attributed to web browsing, which is more 
intermittent than saturated. The reason that it is intermittent rather than
saturated is that TCP involved; we will discuss this in more detail shortly.

While TCP-controlled file downloads in WLANs have been studied in
\cite{astn_model:Kuriakose}, it was assumed that all STAs are associated
 with the AP as a \emph{single} rate. In reality, multiple rates are possible;
for example, 802.11g offers 8 rates of association. The study of TCP-controlled
file transfers with multiple rates of association has received little attention
 in the literature; a first step was taken in \cite{astn_model:Krusheel}.

Even though the literature on association policies is extensive, the vast
 majority of the works considers the saturated traffic model. Given the
prevalence of TCP-controlled traffic, which does not correspond to the 
saturated model, and the fact that multiple rates of STA-AP association are
possible, one begins to wonder if policies based on a different approach
can perform better. Clearly, if this were possible, a very substantial
number of WLAN users would stand to benefit.

This question motivates the work reported in this paper. We obtain an
 association policy for STAs in an IEEE 802.11 WLAN
by taking into account explicitly two aspects of practical importance:
(a) TCP-controlled short file downloads interspersed with read times
(motivated by web browsing), and (b) different STAs associated with
an AP at possibly different rates (depending on distance from the AP).

\begin{figure}
\centering
\includegraphics[scale=0.5]{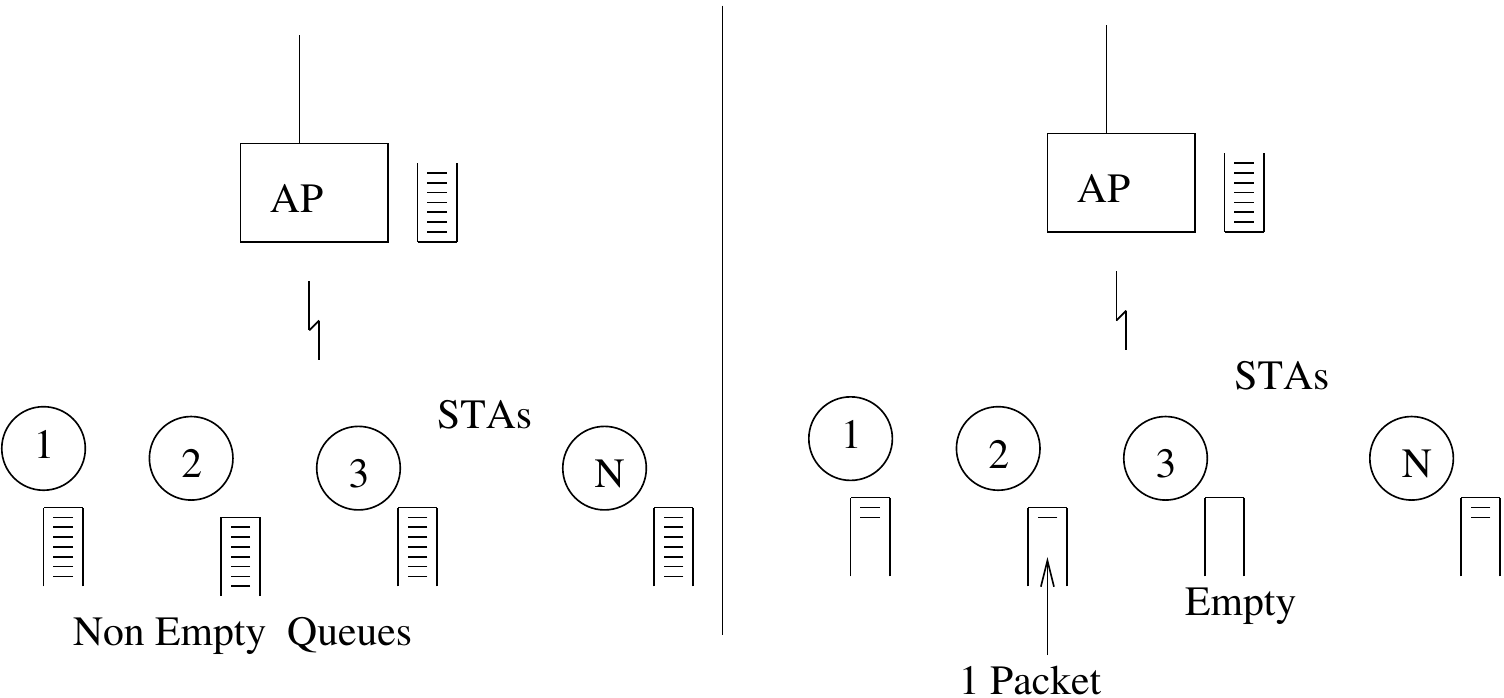} 
\caption{Two scenarios, the left side one with the saturated traffic and the
 right side one with unsaturated traffic. The queues of the AP and STAs are
shown next to them. In the saturated traffic case both the AP and the STAs have
packets in its queue all the time. In the unsaturated traffic scenario which we
are addressing, the AP only have packets to transmits all the time and the MAC
queues at STAs are empty most of the time. }
\label{fig:Traffic_model}
\end{figure}

To see why unsaturated traffic makes a difference, we consider Figure 
\ref{fig:Traffic_model}. The left part shows a saturated traffic scenario,
where all WLAN entities have packets to transmit; therefore, $(N+1)$ entities
contend for the channel. The right part shows the situation with TCP in the
picture. Essentially, for many TCP connections, the entire window of packets
sits at the AP, leaving the corresponding STAs with nothing to send. This means
that the number of contending WLAN entities is much smaller
\cite{astn_model:Kuriakose}. This suggest why approaches relying on saturated
traffic may be inadequate.

\begin{figure}
\centering
\includegraphics[scale=0.45]{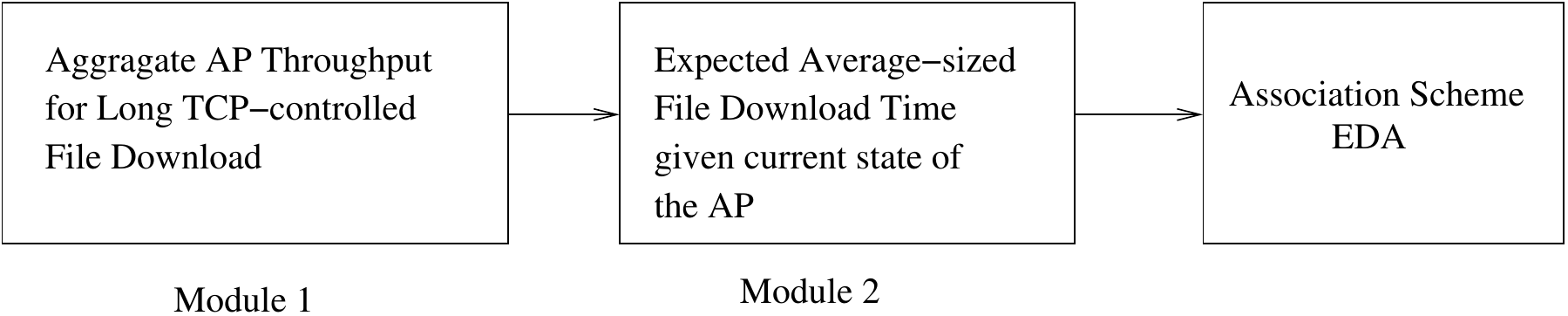} 
\caption{ Approach of the modelling and application of the model}
\label{fig:blockdiagram}
\end{figure}

We approach the problem in two steps. First, we consider an analytical
model to obtain the aggregate AP throughput for long TCP-controlled
file downloads when STAs are associated at $k$ different rates $r_1$,
$r_2$, $\ldots$, $r_k$; this extends earlier work in the literature.
Second, we present a 2-node closed queueing network model to approximate
the expected average-sized file download time for a user who shares the
AP with other users associated at a multiplicity of rates.
These analytical results motivate the proposed association policy,
called the Estimated Delay Based (EDA) policy:
Associate with the AP at which the expected file download
time is the least. To summarize, our approach can be depicted as shown in Figure
 \ref{fig:blockdiagram}.

Our contributions are as follows:\\
 (i) We  derive a closed form expression for the aggregate throughput of an AP
 with which STAs are associated at multiple data rates. Simulations indicate a 
 very good match with analytical results.\\
 (ii) We obtain a closed form expression for the mean delay experienced by an STA
  in downloading an average sized file. Again, simulation results match the analysis 
 very well.\\
 (iii) Utilizing these results, we present an on-line, distributed, 
 client-driven association policy (EDA) which gives substantially increased throughput
 for every STA with balanced load among the APs; throughputs are observed to improve
 by 12.8\% to 46.4\%.

\textbf{Outline of the paper:}
 In Section \ref{sec:Related_Work} we provide a brief overview of related 
work. In Section \ref{sec:System_Model}  we state the assumptions, describe our
models and introduce notations. In Section \ref{sec:Aggr_Thpt}, we analyze a 
model for aggregate throughput of an AP with multi-rate association. Section
\ref{sec:BCMP_network} covers the analysis of average download time in a
multirate scenario. In Section \ref{sec:New_association_policy}, we present a new 
distributed association policy based on the average file download  delay.
In Section \ref{sec:Performance_Evaluation}, we provide simulation results that compare
EDA policy with the SNR-based and aggregate throughput 
maximizing schemes. In Section \ref{sec:Extension} we discuss the results and 
identify future research directions.  
Finally, in Section \ref{sec:Conclusion}, we conclude the paper.
%related work section starts here.
\section{Related Work}\label{sec:Related_Work}

From the extensive literature on association in WLANs, several themes can be 
discerned. In the first group, a newcomer STA assesses each AP 
(that it can hear) by finding out the per-STA throughput received by STAs 
associated with it currently \cite{astn_model:Fukuda1}, \cite{astn_model:Fukuda3} and 
\cite{astn_model:AlRizzo}. In these papers, the authors derive expressions for aggregate AP throughput  (with UDP traffic), taking into account the number of STAs associated, as well
as the probability of packet loss due to collisions. From this, the per-STA
throughput can be calculated, and the newcomer associates with the AP for which
the per-STA throughput is the highest.

In the second group, we have association policies that start from the throughputs
that individual STAs are getting at present \cite{astn_model:Kauffmann}, \cite{astn_model:Abusubaih2}, \cite{astn_model:Ekici1}, \cite{astn_model:Ekici2} and 
\cite{astn_model:Bejerano1}. For each STA associated with an AP, the ratio of 
throughput to the physical rate of association is obtained, and this fraction is
added over all the STAs. This leads to a figure of merit for each AP, and the 
newcomer chooses the AP with the highest figure of merit.

For a brief overview of the other categories, let us assume that a newcomer
STA can hear 3 APs denoted as $a$, $b$ and $c$, with $n_a$, $n_b$ and $n_c$ STAs (respectively)
associated with them. Suppose the new AP associates with AP $a$. Then, one can imagine
an $(n_a +n_b + n_c +1 )$ -element vector $\overline{v _a}$, listing the throughputs of all STAs in
the system. Similarly, there would be vectors corresponding to choosing AP $b$ $ \overline{v_b} $ and AP $c$ $\overline{v_c}$.

In the third group, the selection of the AP is made by considering Jain's fairness
index $JF(\overline{v_a})$, $JF(\overline{v_b})$ and $JF(\overline{v_c})$, corresponding to $v_a$, $v_b$ and $v_c$ 
respectively. The AP selected is the one for which the Jain Fairness Index 
is the highest \cite{astn_model:Jabri}. In the fourth group, $\overline{v_x}$ is 
mapped to an expression for AP load $(x = a,b,c)$, and the AP which is least
loaded is selected. A variation considers a map from the vector of throughputs 
$(\overline{v_x})$ and the vector of rates of association to the load \cite{astn_model:Kasbekar2} . Finally, 
some association policies base their choice on maximizing the minimum 
element of $\overline{v_a}$, $\overline{v_b}$ and $\overline{v_c}$ (``max-min fairness''), or maximizing
the sum of the logarithms of the elements of $\overline{v_x}$, $x = (a, b, c)$, 
\cite{astn_model:Liew}, \cite{astn_model:Gong2}, \cite{astn_model:Siris}, \cite{astn_model:Li}, and \cite{astn_model:Bejerano2}.

A common assumption in all these policies is that of saturated traffic, viz., all WLAN entities have packets to send always.

IEEE 802.11$k$ \cite{astn_model:IEEE} define several metrics for radio resource management, including channel load report, BSS average access delay and BSS load. Any 
STA can request for channel load report. Other load metrics are broadcast 
by AP in beacon frames. These metrics can be utilized in arriving at association decisions.

The proposed EDA policy leverages the capabilities provided by 802.11$k$.

\section{Problem Statement and Models}\label{sec:System_Model}

Our goal is to look for an association policy that takes into account the 
specifics of TCP-controlled file transfers. To move towards this goal, our 
approach is to model two aspects: The aggregate AP throughput achievable in a 
multirate scenario and the expected time to download a file. Models for these 
two aspects are developed in Section \ref{sec:asm_ap} and \ref{sec:asm_web}
 respectively. In Section \ref{sec:asm_gen}, we discuss the general assumptions that 
we have made.

\begin{figure}
\centering
\includegraphics[scale=0.7]{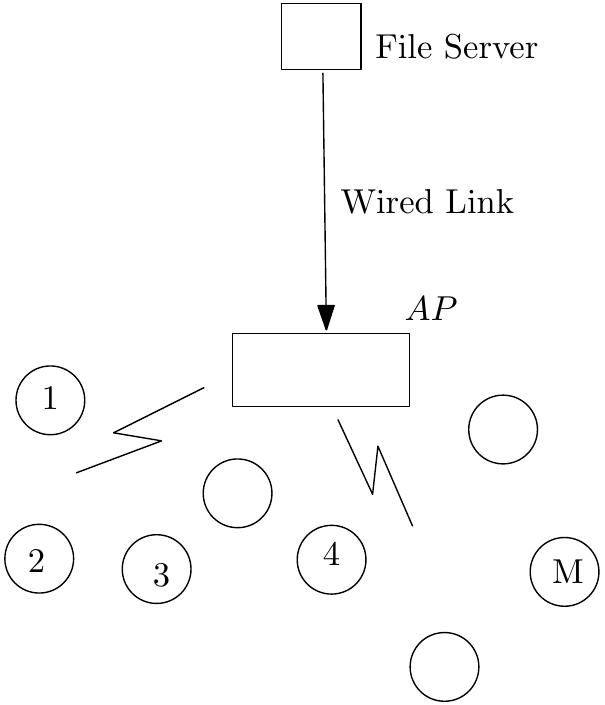} 
\caption{ Fixed number of STAs downloading files from a server which are 
connected through an AP. No new arrival or departure of the 
STA}
\label{fig:AP_STA}
\end{figure}

\subsection{General assumptions}\label{sec:asm_gen}
%subsection III A
We consider a single cell 802.11 a/b/g/n of $M$ (large) STAs associated with a 
single AP as shown in Figure \ref{fig:AP_STA}. All nodes are contending for 
the channel using the DCF mechanism. STAs are associated with the AP at $ k $ 
different physical rates. We assume that there are no link error and
packets in the medium are lost due to collisions only. The AP is 
connected to a file server by a wire as shown in Figure \ref{fig:AP_STA}. In
this paper, we assume that round trip time (RTT) is very small.

\subsection{Model for aggregate AP throughput}\label{sec:asm_ap} 
%subsection III B
Each STA has a single TCP connection to download a large file from a local 
file server. In other words, the traffic model is homogeneous across STAs and
no other traffic is present. The AP delivers TCP packets towards STAs and an
STA returns TCP-ACK packets. There are several simultaneous TCP connections;
every STA and the AP is contending for the channel. Further, as no preference is
given to the AP, the AP is modelled as being backlogged permanently 
\cite{astn_model:Kuriakose}. We assume that the AP uses the RTS-CTS 
mechanism while sending packets to the STAs
and an STA uses basic access to send TCP-ACK packets. As soon as an STA 
receives a data packet, it generates a TCP-ACK packet without any delay and 
this is placed in the MAC queue. Also, we assume that all the nodes have 
sufficiently large buffer space, so that packets are not lost due to buffer 
overflow. Moreover, recovery form collision occurs before TCP time outs and
 TCP start-up transients are ignored.
%subsection III C
\subsection{Model for estimating file download time}\label{sec:asm_web}
We consider a given AP and a fixed number of STAs associated with it; in 
other words, no new STA arrives and no associated STA leaves the system. 
Aggregate throughput of the AP is shared equally among all $ M $ STAs. Hence 
the AP is modelled as a processor sharing queue. We assume that a file downloaded
by an STA can belong to one of $L$ different classes, A file of class $l$, 
$ 1 \leq L $, is distributed exponentially, with mean size $ 1/\mu_l $. After 
downloading a file, an STA goes into the ``reading'' state. 
For a class $l$ file, the reading time is distributed exponentially also, with 
the mean being $1/ \lambda_l $. The reading state is modelled by a ./M/$\infty$ 
queue. Further, upon completion of reading, an STA downloads a fresh file of 
random size having mean $ 1/\mu_l $  with probability $ q_l $, 
$\sum_{l=1}^{L} q_l = 1$. Together, the AP queue and the ``reading'' queue constitute
a closed queueing network.

\begin{figure}%[h]
\centering
\includegraphics[scale=0.7]{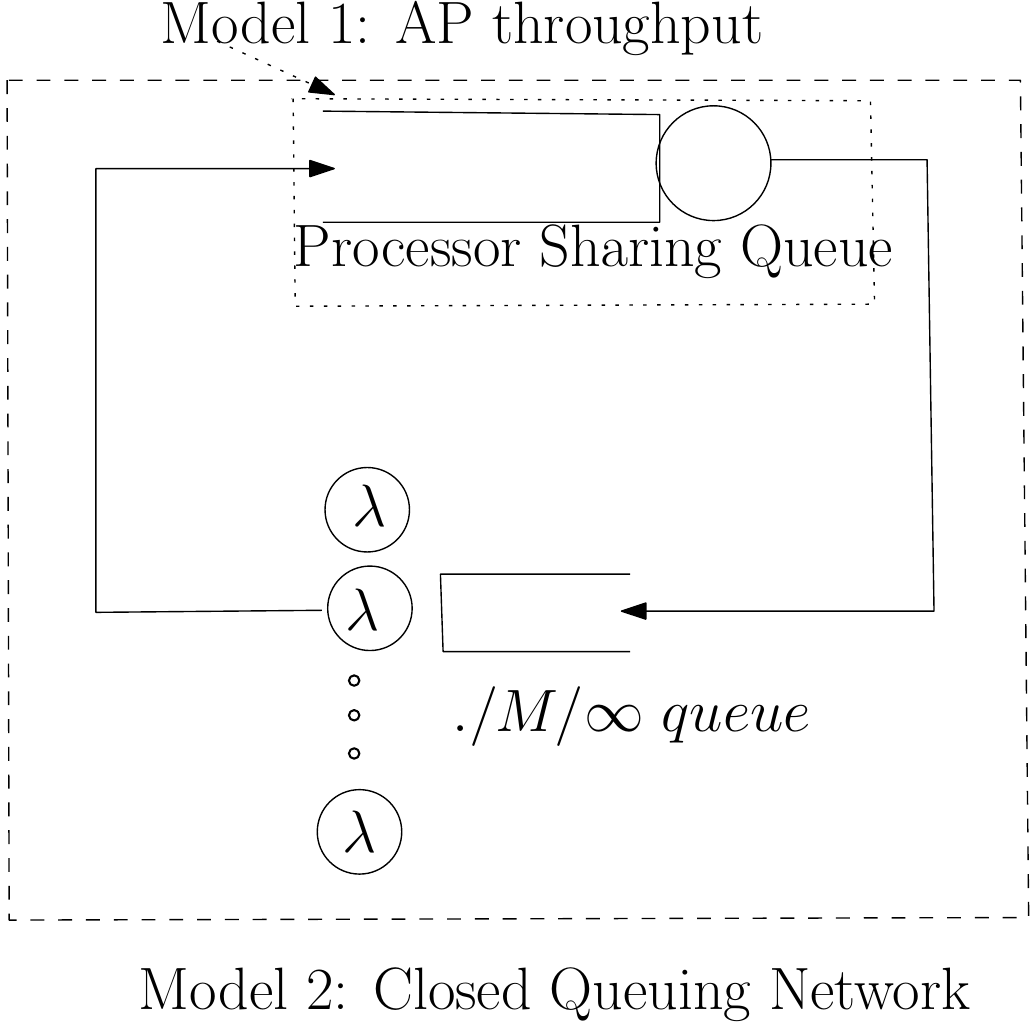}
\caption{A closed queue network model in which constant number of STAs 
alternating between Processor Sharing server (AP) and $ M/M/ \infty 
$ server }
\label{fig:queue_net}
\end{figure}

\section{Aggregate AP Throughput Model}\label{sec:Aggr_Thpt}

In this section, we are interested in obtaining a method to compute the aggregate throughput achieved by the AP. We assume that each STA is downloading an infinitely long file from a server attached to the AP via a wired network (for example, an Ethernet cable). A TCP packet transmitted to \emph{any} STA is considered in obtaining the throughput expression; hence the resulting expression yields the aggregate throughput across all TCP connections. The result obtained here will be used as a part of the model presented in the next section.

Packets for every STA have to go through the AP and every STA replies with a TCP-ACK. Hence, all STAs along with the AP are contending for the channel. The AP does not have any preference over STAs, and it is modelled as being backlogged permanently. In this condition, we can model the performance of the AP as in \cite{astn_model:prdp_jk}. 
 
 The basic approach is to consider consecutive successful transmissions by
the AP or one of the STAs. At these epochs, we consider the number of STAs
at rate $r_i$, $1 \leq i \leq k $, with TCP-ACK packets in them. This 
$k$-element non negative vector, embedded at consecutive successful 
transmission epochs, can be seen to be a Discrete Time Markov Chain (DTMC),
and this DTMC is part of a Markov Renewal sequence defined at the consecutive
success epochs. To obtain the aggregate AP throughput, we appeal to the Renewal
Reward Theorem \cite{astn_model:AKumar}, where a reward of 1 accrues at every
successful transmission by the AP. Owing to lack of space, we provide only 
a sketch of the argument; more details can be found
 in \cite{astn_model:prdp_jk}.
 
Let $M_i$ be the number of STAs associated with the AP at rate $r_i$, 
where $i \in \lbrace 1,2,\dots k \rbrace$. The analysis proceeds by assuming
$ M_i , 1 \leq i \leq k, $ to be large.
\begin{figure}%[h]
\centering
\includegraphics[scale=.3]{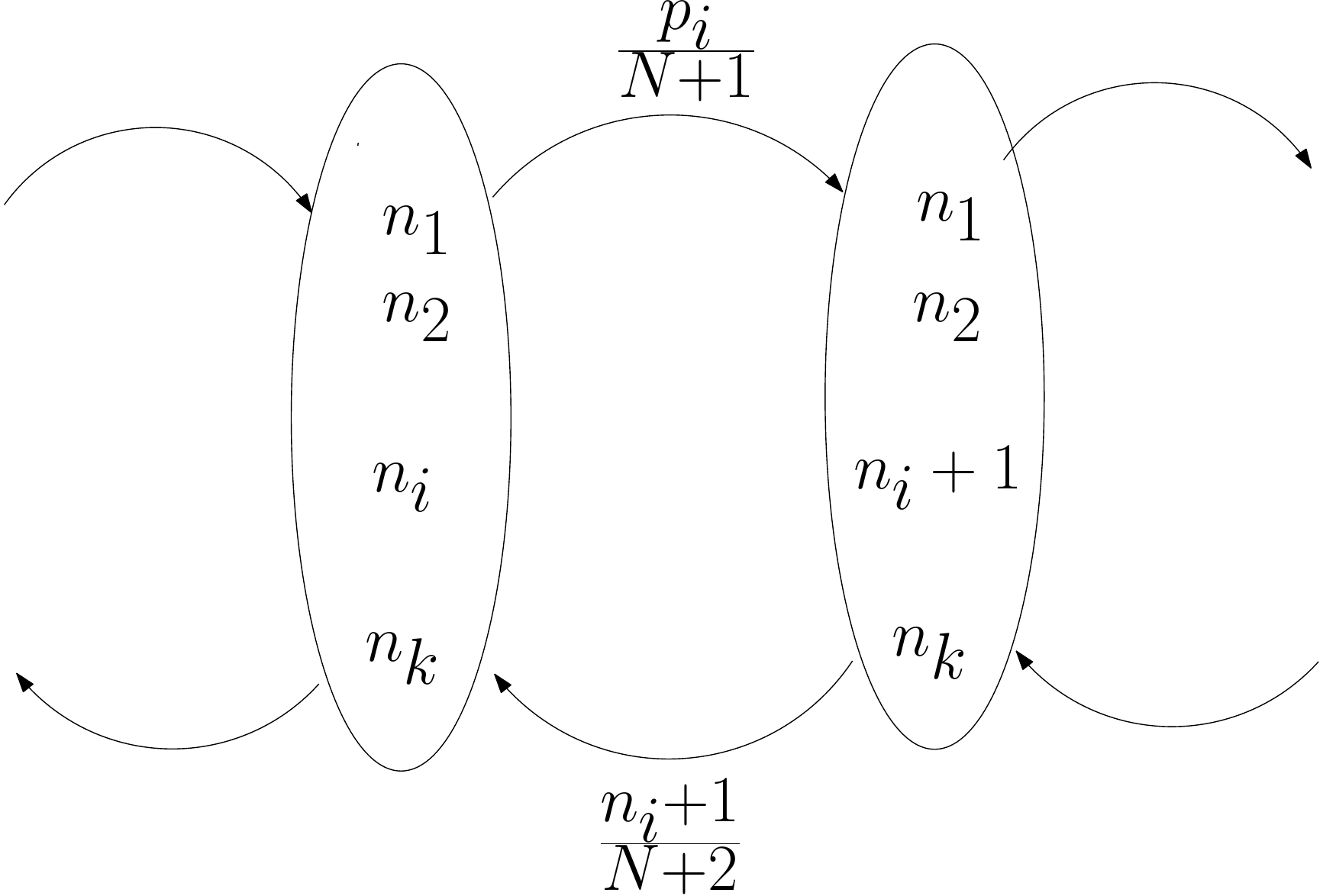}
\caption{Embedded Markov chain formed by an AP and STAs 
associated with the AP at $ k $ different data rates }
\label{fig:MarkovChain}
\end{figure}
As in \cite{astn_model:prdp_jk} we can get the aggregate TCP throughput of
the AP as\\
\begin{equation}	
 \Phi_{AP-TCP} = \frac{\Sigma^{\infty}_{n_1 =0} \Sigma^{\infty}_{n_2 =0} ... 
 \Sigma^{\infty}_{n_k =0} \pi (n_1... n_k)\frac{1}{n_1+...+n_k} 
}{\Sigma^{\infty}_{n_1 =0} \Sigma^{\infty}_{n_2 =0} ... \Sigma^{\infty}_{n_k 
=0}  \pi (n_1... n_k)  E_{n_1,..n_k}X}  
\label{eq:ap_thpt}
\end{equation}

where $ \pi(n_1,n_2,...,n_k) $ is the equilibrium probability of the DTMC being
in state $(n_1,n_2,...,n_k) $ , $X$ is the sojourn time in a DTMC state and 
$E_{n_1,n_2,...,n_k}X $ is the expected sojourn time, given that the DTMC 
state is $(n_1,n_2,...,n_k)$.

To verify the accuracy of  the model, we performed experiments using the Qualnet 4.5 network simulator \cite{astn_model:Qualnet}. We considered 802.11b physical data 
rates as  1Mbps, 2Mbps, 5.5Mbps and 11Mbps, depending on the physical distance from the AP. 
In Table \ref{table:aps_stas}, results are given for a few cases of this 
multirate scenario, i.e., with different number of stations. 

\begin{table}[ht]
\centering % used for centering table
\begin{tabular}{| c |c | c| c | c | c | c | c|} % centered columns (4 columns)
\hline %inserts double horizontal lines
%Case & Method\#1 & Method\#2 & Method\#3 \\ [0.5ex] % inserts table
Total no. & \multicolumn{4}{|c|}{No. of STAs with rate} & \multicolumn{3}{|c|}{ Aggregate Throughput } \\
\cline{2-8} 
of STAs & 11 & 5.5 & 2  & 1 &  Analysis & Simulation & Error \%  \\  [0.5ex]
%heading
\hline % inserts single horizontal line
\multirow{2}{*}{10} & 2 & 3 & 2 & 3 & 2.54 & 2.52 & 0.7 \\ % body of the table
& 1 & 2 & 3 & 4 & 2.57 & 2.53 & 1.6 \\ \hline
\multirow{2}{*}{12} &  2 & 2 & 4 & 4 & 3.49 & 3.45 & 1.2 \\
& 4 & 4 & 2 & 2 & 2.06 & 2.01 & 2.5 \\ [1ex] % [1ex] adds vertical space
\hline %inserts single line
\end{tabular}
\caption{Analysis and Simulation Results [\textit{Mbps}] for multirate AP in 
IEEE 802.11\textit{b} }
 % title of Table
\label{table:ap_model} % is used to refer this table in the text
\end{table}
Detailed analysis is provided in \cite{astn_model:prdp_jk}. From Table 
\ref{table:ap_model}, we see that the analytical approach yields results that
are very close to those obtained from simulations. In the next section, the 
method to calculate aggregate AP throughput will be used to parameterize the 
model that will be presented.

\section{Short File download delay model}\label{sec:BCMP_network}
In this section, our goal is to obtain a figure of merit for a particular AP;
the figure of merit reflects the performance that can be expected if a newly
arrived STA associates with this AP, given its current ``association state''
(that is, for each rate $r_1, r_2, ..., r_k$, the number of STAs associated
with the AP at present). The question we address in this section is: If the 
current association state of the AP remains unchanged for a long time, what
is the performance that a newly arrived STA can expect upon associating with
this AP?

To recall our model in Section \ref{sec:asm_web}: Each user alternates between
downloading a finite-sized file and reading it. The downloaded files belong to
$L$ classes; all files are distributed exponentially, and each class 
corresponds to a distinct mean file size. When a user downloads, the file is 
of class $l$ with probability $q_l$; at each download, the file class is chosen
independently. This behaviour applies to each user; in other words, use
homogeneity is assumed. 

We consider a 2-node closed queueing network with the number of customers equal
of the total number of STAs associated with the AP (Figure \ref{fig:queue_net}).
Node 1 represents the AP, and node 2 represents the ``reading station''. Node 1
is a Processor Sharing service station, because the AP may be considered to be
sharing its aggregate throughput among the STAs that are downloading actively.
Node 2 is modelled as an infinite server service station; it captures the
reading times spent in reading files that have been downloaded already.  At a
given instant, the number of customers at node 1 is equal to the number of STAs
downloading files at that instant. The number of customers at node 2 is equal to
the number of STAs that are reading at that instant. The service rate of the
server at  node 1 is set to the aggregate AP throughput obtained in Section
\ref{sec:System_Model}. The service rates of the servers at node 2 depend on 
the customer class.

We remark upon two key modelling assumptions that we make in this section.
The first is that all wireless-specific aspects are encapsulated in the service
rate of node 1. Thus, the details of the number of STAs associated at the rates
$ r_1 $, $ 1 \leq i \leq k $, influence the service rate of node 1; nothing 
else in the 2-node model is specific to a wireless context. The second 
modelling assumption is that even though the aggregate AP throughput was 
obtained (in Section \ref{sec:Aggr_Thpt}) for a scenario where infinitely large
 files were being downloaded, the same expression is used in the current 
 context, where finite-sized files are being downloaded. Both assumptions are
  made for tractability and simplicity. Not making these assumptions lead to
   much more complex models, and it is not clear that these can be analyzed; 
   even if analytical treatment is possible, it is unclear whether the 
   improvements in the results would justify the complexity. 

Ultimately, the figure of merit that we wish to compute is the average time taken to download a file, given the current association state of the AP.

Let the service rate of the AP be $ \tau $. Let us consider $ m_1 $ STAs to be 
at node 1. That is, $ m_1 $ among $ M $ STAs are busy in downloading files. 
These STAs can download files with mean size $ \frac{1}{\mu_1} $ or 
$ \frac{1}{\mu_2} $ ... $ \frac{1}{\mu_L} $. Now $ m_2 $ out of $ M $ STAs are in
read state. The state of the network can be represented by $ S = ( x_1, x_2 ) $.
Here,  $ x_{i} $ is a vector with elements $m_{i,l} $; where $m_{i,l}$ is the number of customer of class $l$ at node $i$, (index $i$ is for node
number and $l$ for class), $i \in \lbrace 1,2 \rbrace $, $l \in \lbrace1,2,\dots L \rbrace $. Also $m_{i} = \sum_{l=1}^{L} m_{i,l}$.

At node 1, the STA chooses the class $l$ with probability $q_l$, irrespective
of its previous class. But, at node 2, an STA stays in the same class $l$, which
 it had chosen at node 1.
Let $ v_{i_1,l_1,i_2,l_2}$  be the probability that a class $l_1$ STA at node
 $i_1$ becomes a class $l_2$ STA at node $i_2$. Then we have:
$\begin{array}{c l}
\mbox{$ v_{1,l,2,l}  =  1 $ } & \mbox{ $ l \in \lbrace 1, 2,..., L \rbrace $}\\
\mbox{$ v_{1,l,2,j}  =  0 $ } & \mbox{ $ j \neq l $} \\
\mbox{$ v_{2,l',1,l} =  q_l$} & \mbox{ $ (l,l') \in \lbrace 1, 2, ... L \rbrace  $ } 
\end{array} $
%end of equations
For every $i$, let $e_{i,l}$ be the fraction of transitions into class $l$ center $i$.
 Then we have: 
\begin{equation}
e_{i,l} = \sum_{(i',l')} (e_{i',l'}) v_{i',l',i,l}
\end{equation}
After solving the above equations,
\begin{equation}
e_{1,l} = \frac{q_l}{q_1} e_{1,1} = e_{2,l}
\end{equation}
$ e_{i,l}$ is determined to within a multiplicative constant. $ e_{i,l} $ can be interpreted as the relative arrival rate of class $l$ customers to service center $i$.

This is a 2-node BCMP network \cite{astn_model:bcmp}. The processor sharing server in Figure \ref{fig:queue_net} is a type 2 server, and the ./M/$\infty$ is a type 3 server in
the terminology of \cite{astn_model:bcmp}.
The expression for the service rate of node 1 is given by $\Phi_{AP-TCP} $
in \eqref{eq:ap_thpt}, and that of the node 2 server is $ 1/ \lambda_l $
for class $l$ customers. By the BCMP theorem \cite{astn_model:bcmp}, the equilibrium probabilities are given by 
\begin{equation}
P(S=x_1,x_2) = C d(S)f_1 (x_1)f_2 (x_2)
\label{eq:bcmp_dist}
\end{equation}
where $C$ is the normalizing constant chosen to make the equilibrium state probabilities sum to 1. $d(S)$ is a function of the number of customers in the system, and $f_i$ is a function that depends on the type of service center $i$.

From \cite{astn_model:bcmp}, for the Processor Sharing Server, Node 1,
\begin{equation}
f_1(x_1) = m_1! \Pi _{l=1}^{l=L} \frac{1}{m_{1,l} !} (\frac{e_{1,l}}{\tau })^{m_{1,l}}
\end{equation}

and for the infinite server, Node 2,
\begin{equation}
f_2(x_2) = \Pi _{l=1}^{l=L} \frac{1}{m_{2,l} !} (\frac{e_{2,l}}{\lambda_l})
\end{equation}

For a closed network, $ d(S) = 1 $.
 
The average number of  active STAs, $ n_{ avg DS } $ and average number of reading stations $ n_{ avg RS } $  can be obtained by finding the marginal distributions from \eqref{eq:bcmp_dist}. 
 
\begin{equation}
n_{avg DS} = \sum_{ l \in \lbrace 1,2,... L \rbrace } m_{1,l} P(m_{1,l})
\end{equation}

\begin{figure}
\centering
\includegraphics[scale=0.7]{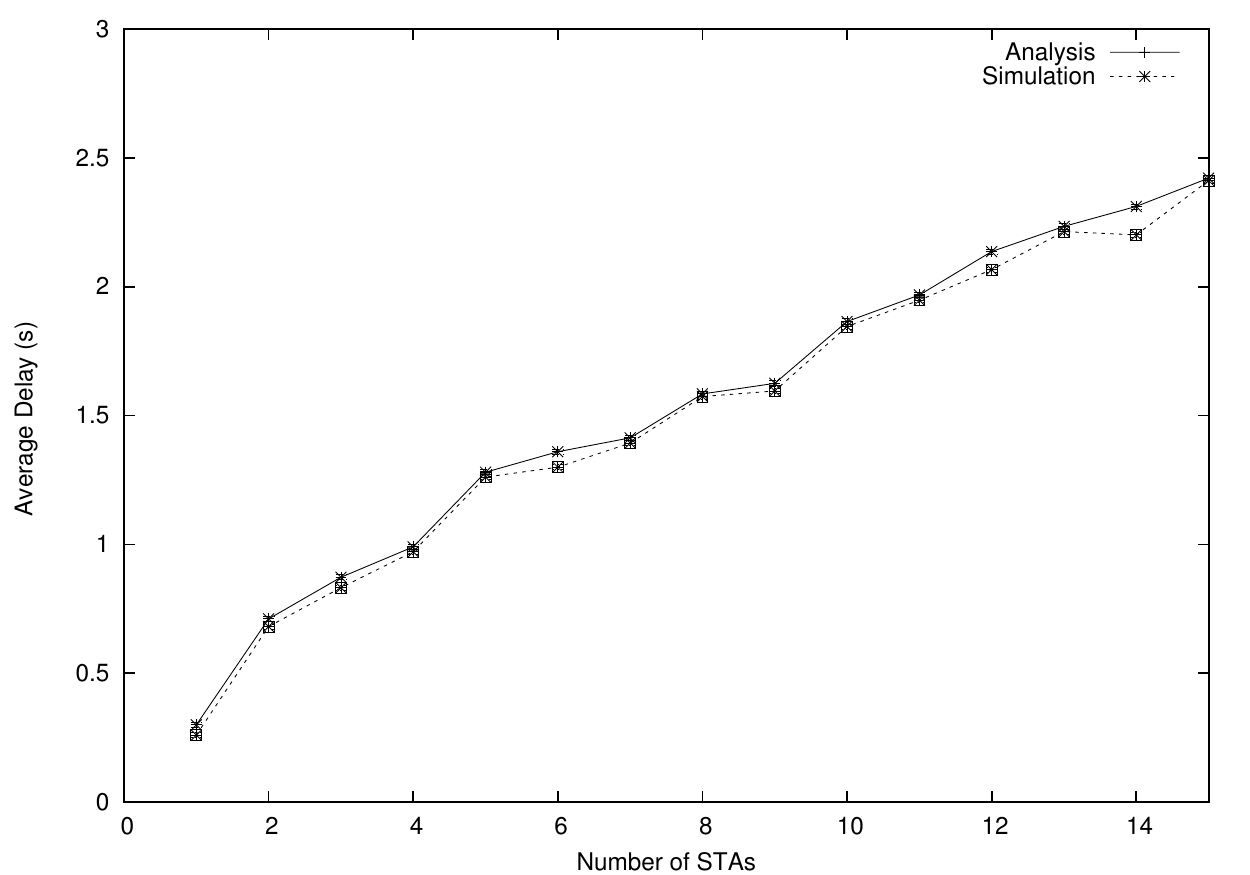}
\caption{ Average delay seen by STAs for downloading exponentially distributed
files with mean 200KB with read time of mean 90 seconds}
\label{fig:Delay}
\end{figure}

From Figure \ref{fig:queue_net} it is clear that  

\begin{equation*}
 n_{avg DS}  + n_{avg RS}  = M  \\
\end{equation*}
Let the throughput in the closed network of Figure \ref{fig:queue_net} be $t_H$.
Then, applying Little's Theorem to node 2, we have

\begin{equation}
n_{avg RS} = t_H \times \sum_{l=1}^{L} q_l \frac{1}{\lambda_l}
\end{equation}
Having obtained $ t_H $ and $ n_{avg DS} $ as above, we can calculate	
the average download delay \textit{d} seen by STAs: 
\begin{equation}
 \textit{d} =   \frac{ n_{avg DS} } { t_H }
\label{eq:est_delay}
\end{equation}
In Figure \ref{fig:Delay}, we compare the results from analysis and simulation
for a single cutomer class. The match is quite good.
% Figure begin 
%end of the figure.
%table begin
\begin{table}[h]
\centering % used for centering table
\begin{tabular}{|c | c | c | c | c | c | c |} % centered columns (4 columns)
\hline %inserts double horizontal lines
%Case & Method\#1 & Method\#2 & Method\#3 \\ [0.5ex] % inserts table
\multicolumn{4}{|c|}{Number of STAs with rate [Mbps]} & \multicolumn{3}{|c|}{Delay [s]} \\ \hline
 11 & 5.5 & 2 & 1 &  Analytical & Simulation & Error(\%)\\  [0.5ex]
%heading
\hline \hline % inserts single horizontal line
1 & 2 & 3 & 4 & 1.954 & 1.938  & 0.82  \\ % inserting body of the table
1 & 3 & 2 & 4 & 1.977 & 1.946  & 1.59 \\
3 & 2 & 3 & 4 & 2.237 & 2.209  & 1.27 \\
2 & 4 & 4 & 3 & 2.352 & 2.314  & 1.64 \\ 
3 & 2 & 4 & 4 & 2.423 & 2.411  & 0.91 \\ [1ex] % [1ex] adds vertical space
\hline %inserts single line
\end{tabular}
\caption{ Comparison of analysis and simulation results for ``delay model''. } 
\label{table:aps_stas} 
\end{table}
Further, to verify the accuracy of the model, we considered a scenario with
fixed number of STAs at different rates. Two different average file sizes with
 mean as 50KB ($ 1/ \mu_1 $) and 250KB ($ 1/ \mu_1 $) represent two different classes in our
model. Any STA selects an exponentially distributed file of mean size 50KB with 
probability 0.6 ($q$) and selects the other type with probability 0.4 ($1-q$). After
downloading the file of mean size $ 1/ \mu_1 $, an STA goes to read mode with
an exponentially distributed duration of mean 25 seconds ($1/ \lambda _1$). If the STA
downloads random file with the other mean (i.e., $1/ \mu_2$), it goes to read mode
with random time of mean 100 seconds ($1/ \lambda_2 $). Finally, after completion 
of read time, STA forgets what it has chosen as mean size
and selects a fresh file from the two classes with probabilities $ q $ and
$ 1-q $. The results are presented in Table \ref{table:aps_stas}. The analytical
values match the simulation results very well.
% Section :association policy 
\section{A new association policy}\label{sec:New_association_policy}
The basic idea of our association policy is simple: For each AP that the newly
arrived STA hears from, a figure of merit (mean download delay) is computed as in the last section.
Then, the decision is to associate with the AP with the best figure of merit.

To calculate the mean download delay, any STA needs to know only the number of STAs associated
with the AP and their physical data rates. This ``association state'' can be 
obtained using IEEE 802.11$k$. If the mean file sizes are known to the STA, then it can calculate the expected delay \textit{d} itself. We assume that
the number of file classes $L$ and corresponding mean file sizes $ 1/ \mu_l$,
$ 1 \leq l \leq L $, are available from offline measurements, from a repository like \cite{astn_model:surge}.
% algorithms  section stats 
\subsection{Algorithm}
The following algorithm runs in every STA as soon as it arrives 
in the network 
\begin{algorithmic}[1]
\STATE Set Channel index, $ c = 0 $ 
\STATE Set AP index, $ j =0 $
\IF { Scan Mode = Active}
\STATE Send Probe Request frame
\STATE Receive the Probe Response frame
\ELSE
\STATE Wait for Beacon frame  
\ENDIF
\STATE Measure the signal strength and find out possible data rate
\STATE Extract the $M_i$ and $ r_i $ from received frame
\STATE Calculated the expected delay $ d_{j} $ for APs $ j $ given 
by Equation \ref{eq:est_delay}\\
\STATE Increment the AP index $ j = j + 1 $
\STATE Goto step number 5 for information from next AP on the same channel.
\STATE Increment the Channel index $ c = c + 1  $ and Goto step number 3 to
 scan the next channel
\STATE Find out minimum of $ d_{j} $ for all $ j $ and across all the channels
\STATE Send Association request to that AP.
\end{algorithmic}
% algorithms  section ends 
\section{Performance Evaluation}\label{sec:Performance_Evaluation}
We present simulation results that evaluate the AP 
selection approach based on the average delay (EDA) discussed in the previous 
section. In particular, we compare the EDA AP 
selection scheme with the RSSI based selection scheme and other similar selection 
schemes.

\subsection{Simulation setup}
We have simulated association policies in the Qualnet 4.5 simulator.
 We enhanced the 802.11 MAC implementation in Qualnet 4.5. Small changes,
 like appending of extra information in beacon frames and probe
frames, have been implemented in the AP module of the simulator. The AP keeps
track of the number of STAs associated with it and the rates of association of 
those STAs. When transmitting beacon frames, it adds this extra information 
to the existing frame and broadcasts it. To handle the association policy in 
active scanning, the AP sends the same information in the probe response frame.
We have implemented the algorithms explained in previous sections and 
corresponding changes required for the association schemes in the STA module.
%subsection to explain scenarios
\subsection{Scenarios}
\begin{figure}
\centering
\includegraphics[scale=.4]{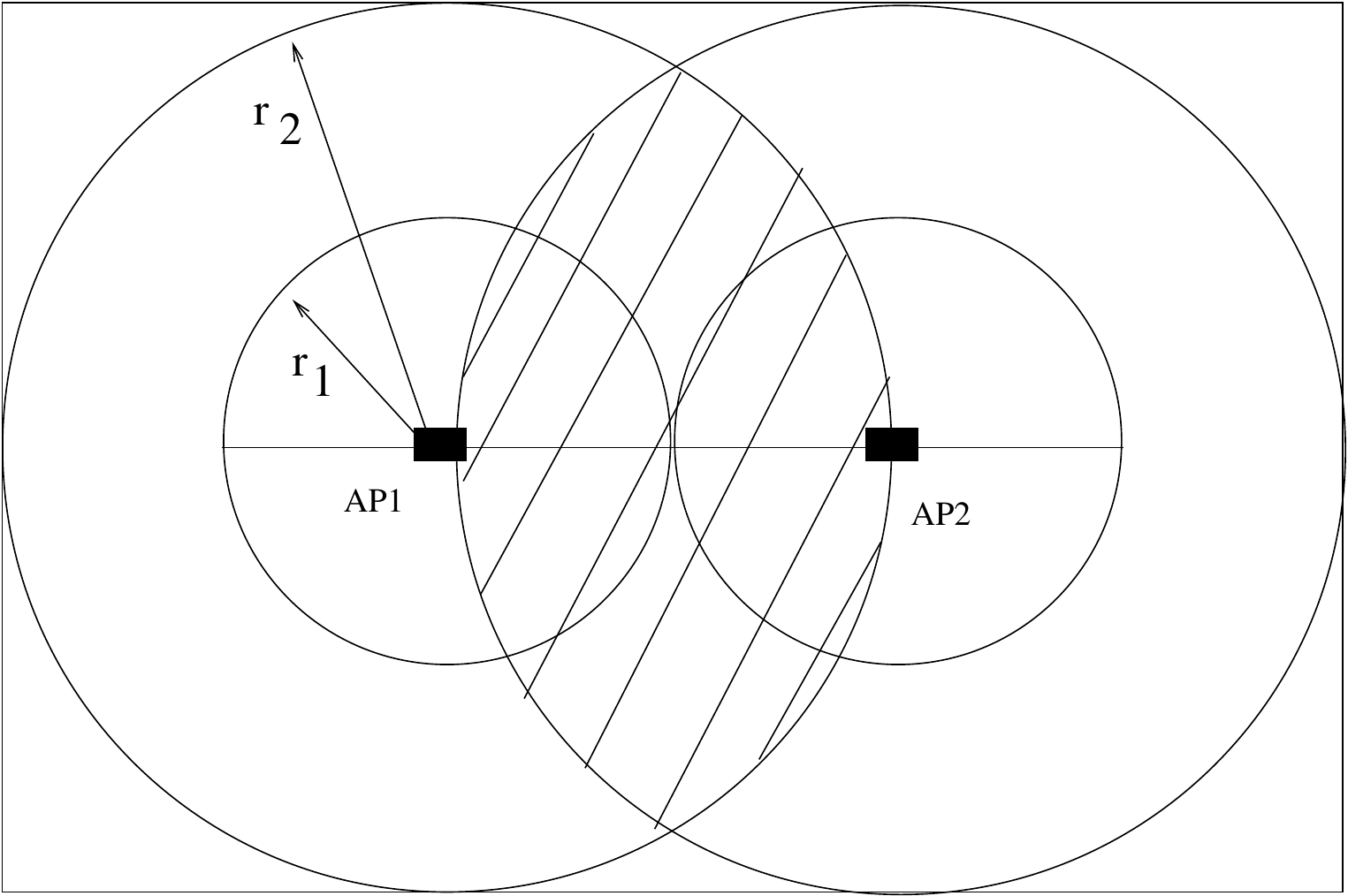}
\caption{$ P_{centre} $ is the fraction of STA arrivals to the shaded region
for the 2 APs scenario. Distances for only two data rates are shown. The outer one
is the maximum distance for lowest data rate. The inner circle is the maximum distance for the highest data rate}
\label{fig:ap2_shade}
\end{figure}

\begin{figure}
\centering
\includegraphics[scale=.5]{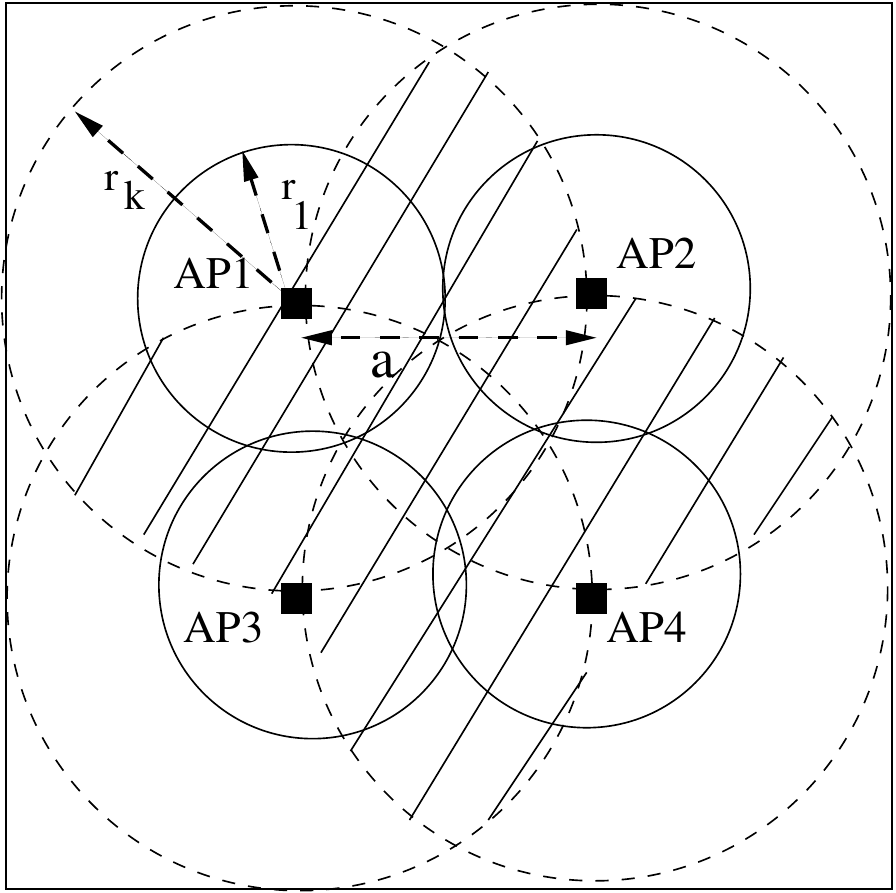}
\caption{Scenario with 4 APs and ranges shown with dotted circles are highest and 
lowest physical transmission rate. Shaded region shows $P_{centre}$ }
\label{fig:ap4_shade}
\end{figure}

\begin{figure}
\centering
\includegraphics[scale=.5]{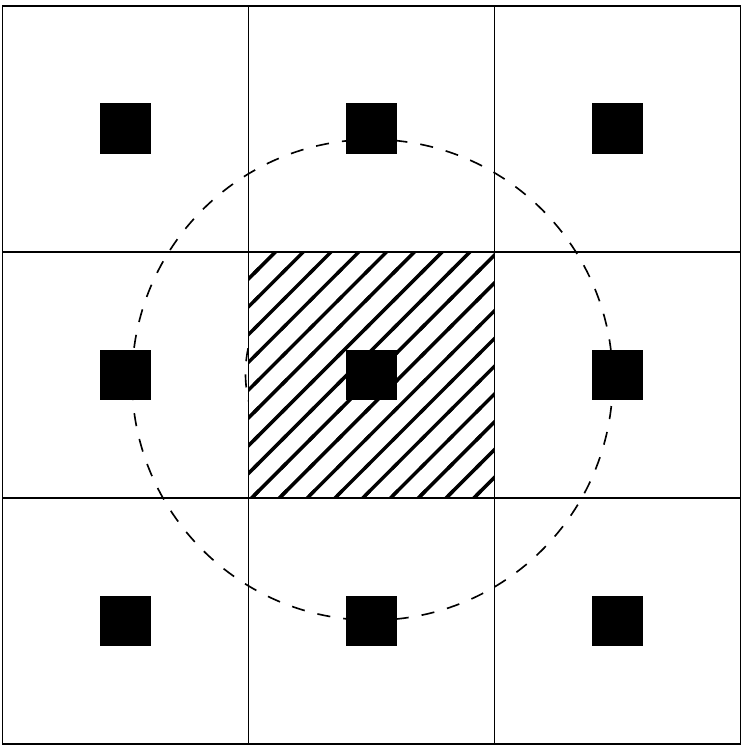}
\caption{Scenario which shows the placements of 9 APs. Shaded region is the
$ P_{centre} $, fraction of overall arrival of STAs to complete region }
\label{fig:ap9_shade}
\end{figure}

We have used three scenarios to understand the performance of the proposed
association scheme. In the first scenario shown in Figure \ref{fig:ap2_shade},
we placed 2 APs at a distance of 480 meters. So, an STA 
very near to one AP can hear the second AP's signal with SNR corresponding to
the lowest possible rate. STAs arrive according to a Poisson distribution with 
parameter $ \nu $ to the region shown in Figure \ref{fig:ap2_shade} which is served by 2 APs. 
STAs which arrive to the shaded region choose between APs; other STAs have no
choice. Hence we considered spatial arrival of STAs to be non 
homogeneous. This is modelled using the parameter $ P_{centre} $, which is the 
probability that an STA arrives within the shaded region shown in Figure 
\ref{fig:ap2_shade}. This is a parameter that can be varied and system 
performance can be studied. After arriving, every STA downloads a file; after
 completion of download, it goes to the read state and then again downloads 
next file. There are two classes of files ($L$=2). The file sizes are distributed
exponentially with means 50KB and 750KB. Read times are distributed exponentially 
with means of 40 seconds and 120 seconds. After downloading a random number of files with 
mean 100, the STA departs from the BSS. We ran a large number of independent 
simulations with different seeds to obtain confidence intervals.
%Starting of Figure
%Ending of Figure
To verify and evaluate the scheme further, in the next simulation 4 APs are
placed as shown in Figure \ref{fig:ap4_shade}. Here again the shaded region is 
the $P_{centre}$ fraction of complete area. This parameter is varied from
0.1 to 0.9 to study the system performance.
%Start figure for 4 AP scenario
% End of figure for 4 AP scenario
And finally we took a scenario with 9 APs as shown in Figure 
\ref{fig:ap9_shade}. Placement of APs are done by considering radio range of 
the APs and STAs. Here in this case, to study the performance of system, the 
load of centre AP is varied by parameter $p_{center} $ as the probability of 
STAs which arrives to the shaded region. Distance between the access 
points are chosen such that any STA in shaded region will have option to 
choose between at least 2 APs. There is always a chance for neighbour APs to share the 
load of centre AP.  
%Starting Figure for 9 AP
%Ending Figure for 9 AP
\subsection{Other Association Policies}
This section focuses on how the proposed EDA policy 
performs in comparison with the SNR based scheme and other 
distributed association schemes.

A distributed association policy based on Available Admission Capacity (AAC)
is presented in \cite{astn_model:Lee}. AAC is the fraction of time of an AP 
that a new STA can acquire. A new association metric based on this EVA
(Estimated aVailable bAndwidth) is proposed. From this metric, an STA can 
predict the maximum achievable rate that it can get. This new metric 
considers multirate, greedy users.
\cite{astn_model:Garcia2} and \cite{astn_model:Gong1} propose association 
polices similar to this AAC based one.

To calculate AAC, traffic offered is calculated and the ratio of this with
 maximum possible traffic is found. The Ratio is used to determine the best AP.
Similarly, to find EVA the expected back off interval is obtained.
 Then channel busy time and contention overhead are calculated. 
 This information is used to select the AP.

We have chosen another client driven association policy for comparison.
The number of associated STAs and the effect of loss of a packet is considered in 
 \cite{astn_model:Fukuda1} and \cite{astn_model:Fukuda3}, while proposing 
an association scheme. The load of the AP is defined by using these two parameters as,
 $(1-P_{error})/n$, where $P_{error} = c_1 (RSSI c_2 + c_3 )^{c_4} $ where $c_1$, 
$c_2$, $c_3$ and $c_4$ are constants, and $n$ is the number of STAs. The APs 
pass these information to STAs in beacon frames.

\subsection{Comparison}
We have implemented the above algorithm in the Qualnet 4.5 simulator.
We used similar scenarios for comparison. 
Table \ref{table:Compare_thpt} shows the comparison of other 
selected policies with our policy.
Let $ t_u $ be the time download time for a file of size $m_u$ and $L$ be the number
of file transfers. Then $ Th_{ij} = \frac{1}{L} \sum _{u=1} ^{L} \frac{m_u}{t_u} $ 
is throughput of STA $i$ which is associated with  AP $j$. The 
average throughput of all the APs is calculated by taking the 
sum of the throughput for each STA, which is given by:\\
\begin{equation}
Th_{avg} = \frac{1}{a}\sum_{j} \sum _{i} Th_{ij}
\end{equation}
Where $a$ is the total number of APs in the network. 
%%%%%%%%%%%%%%%
\begin{table}[h]
\centering % used for centering table
\begin{tabular}{| c | c | c | c | c |} % centered columns (4 columns)
\hline %inserts double horizontal lines
%Case & Method\#1 & Method\#2 & Method\#3 \\ [0.5ex] % inserts table
Scenario & EDA & Load Balancing  & EVA & SNR based \\  [0.5ex]
%heading
\hline \hline % inserts single horizontal line
2 AP        & 2.95$\pm$0.01  & 2.68$\pm$ 0.01 & 2.78$\pm$0.01 &  2.45$\pm$0.01 \\ \hline
improvement &                & 10\%           &   6 \%        &  18.3 \% \\ \hline
4 AP        & 3.80$\pm$0.01  & 3.09$\pm$ 0.01 & 3.45$\pm$0.01 &  2.84$\pm$0.01 \\ \hline
improvement &                & 22.7\%         &  10.1\%       & 33.4\%   \\ \hline   
9 AP        & 4.12$\pm$0.01  & 3.20$\pm$ 0.01 & 3.65$\pm$0.01 &  2.81$\pm$0.01 \\ \hline
improvement &                & 28.7\%        &  12.8\%       &   46.4\%     \\  
\hline %inserts single line
\end{tabular}
\caption{ Comparison of EDA with other policies} Throughput [MBps] with 95\%
confidence intervals obtained by simulation for 3 different scenarios shown in 
Figures \ref{fig:ap2_shade}, \ref{fig:ap4_shade} and \ref{fig:ap9_shade}. 
$P_{centre}$ = 0.9, average file sizes of 50KB and 750KB with reading times as
40s and 120s respectively. Probability of selecting smaller file is 0.6. Arrival
rate of STA is 0.5 per second.
\label{table:Compare_thpt} 
\end{table}
%%%%%%%%End of Table.

\begin{figure}
\centering
\includegraphics[scale=.7]{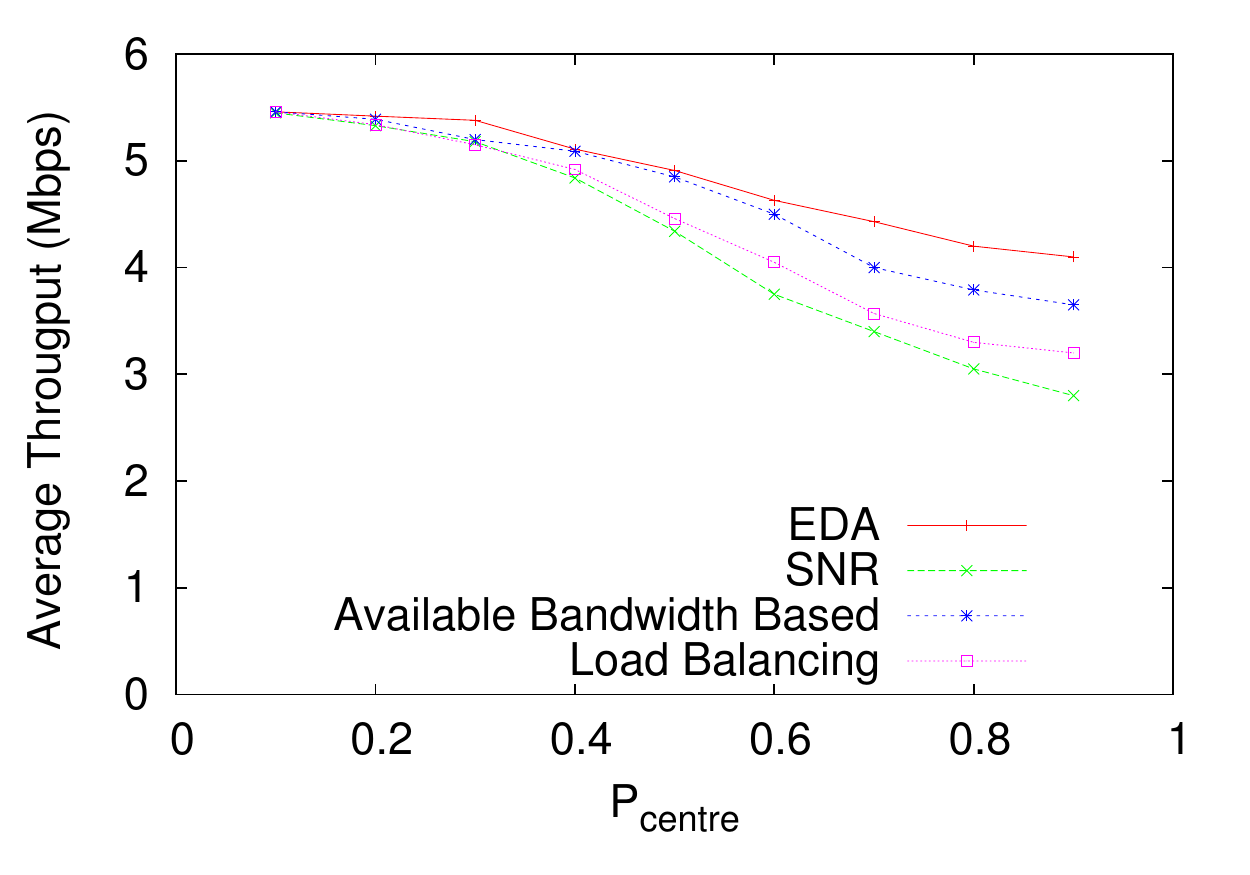}
\caption{Throughput vs. $P_{centre} $  with arrival rate of STAs as 0.5 
per second in the scenario of Figure \ref{fig:ap9_shade} with 9 APs}
\label{fig:Confidence_05}
\end{figure}

\begin{figure}
\centering
\includegraphics[scale=.7]{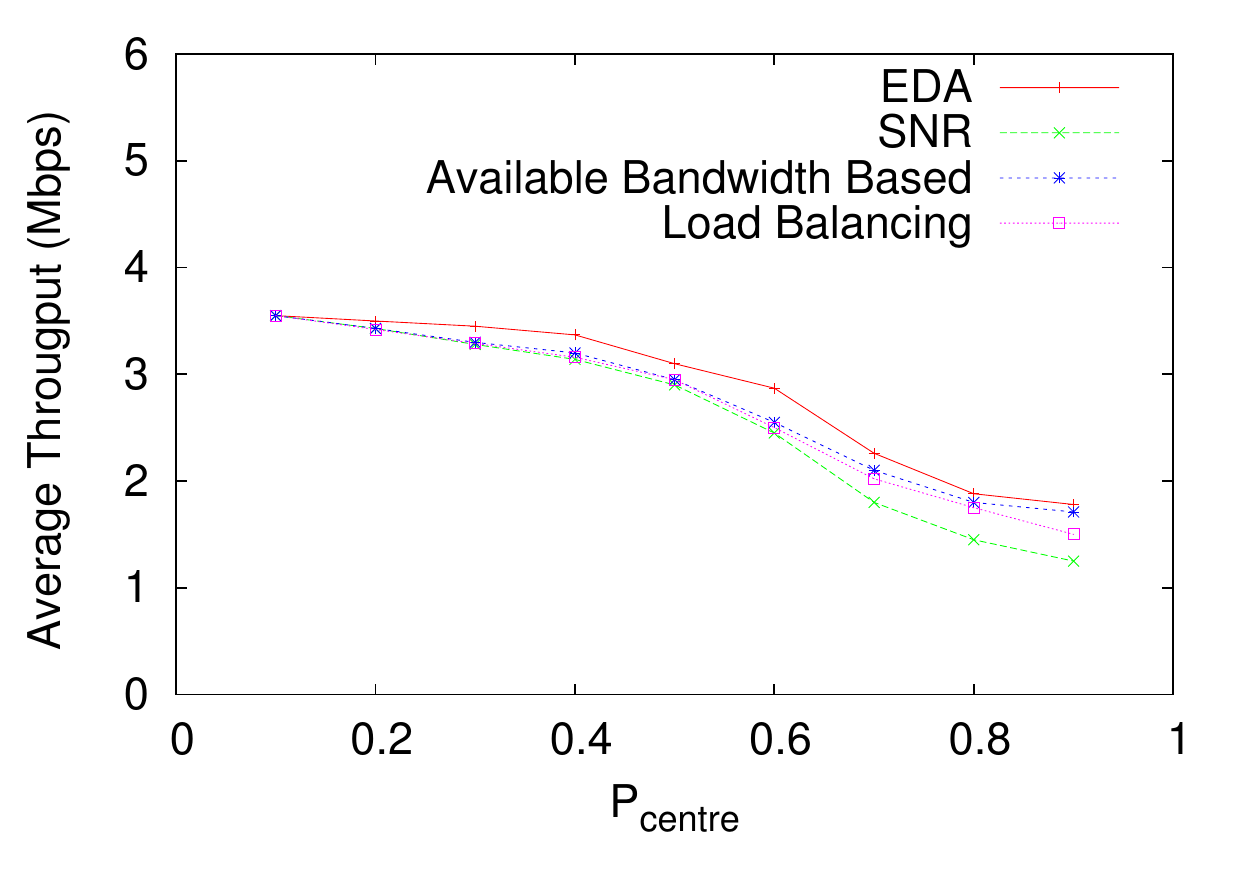}
\caption{Throughput vs. $P_{centre} $  with arrival rate of STAs as 1 per 
second for scenario shown in Figure \ref{fig:ap9_shade} 		}
\label{fig:Confidence_1ps}
\end{figure}

The graph shown in Figure \ref{fig:Confidence_05} is the average of 50 runs 
of the same experiment with the same parameters with different seeds. The arrival
 rate is chosen as 0.5 per second. The 95 \% 
 confidence interval is also shown in the graphs for the aggregate throughput. 
Similarly for the graph in Figure \ref{fig:Confidence_1ps}, the arrival rate STAs 
is 1 per second.
% graph
%graph
%end graph

It can be seen that EDA performs distinctly better than the other policies.
From Figure \ref{fig:Confidence_05}, we see that the extent of improvement
increases as the arrival process becomes more skewed spatially (higher values
of $ P_{centre} $ ). The relative improvement is not as appreciable in
Figure \cite{fig:Confidence_1ps}, all APs are subjected to higher load, and different policies cease to make a significant difference.

\begin{figure}
\centering
\includegraphics[scale=.7]{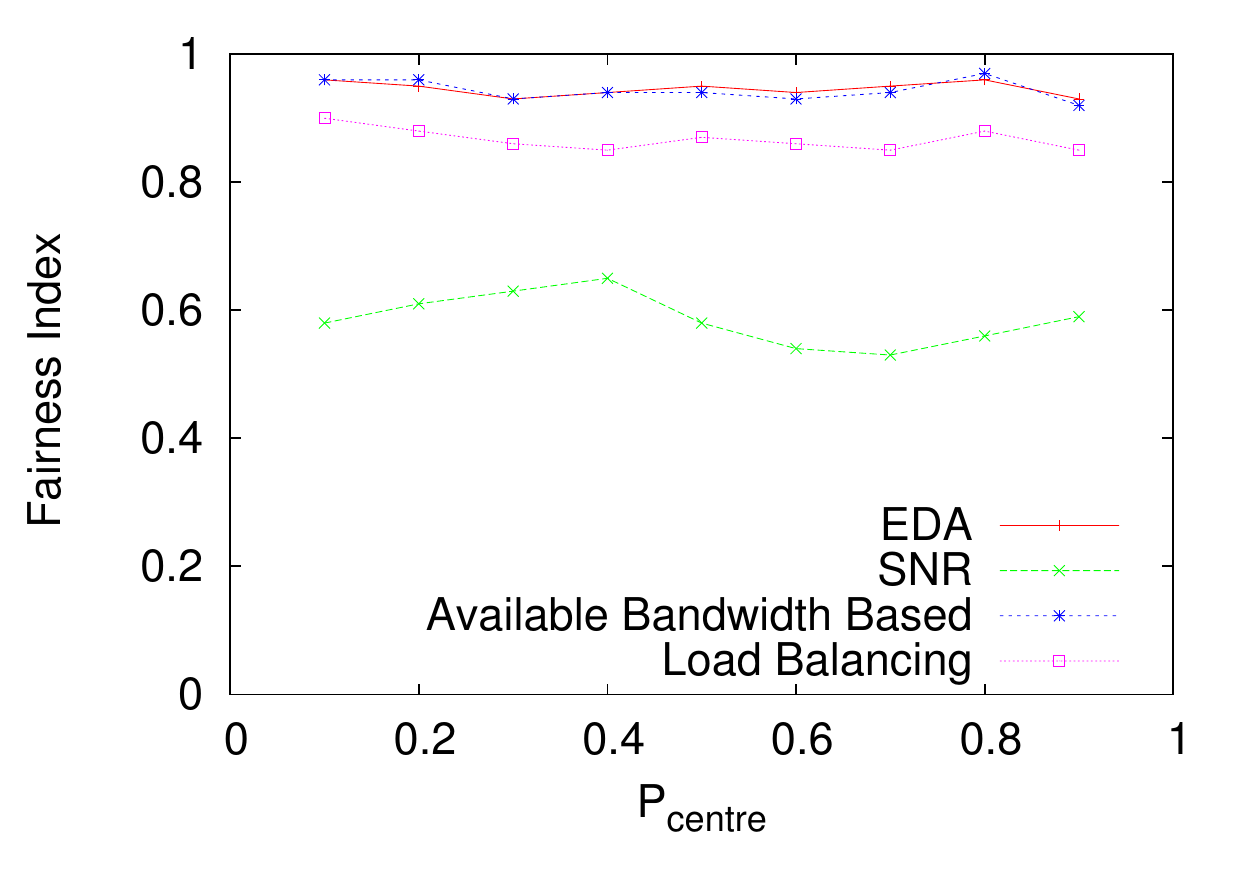}
\caption{Fairness vs. $ P_{centre} $ shows how the fairness index varies with 
increasing the load of centre AP in Figure \ref{fig:ap9_shade} }
\label{fig:fairness}
\end{figure}

 Next we compare the four AP selection methods in terms of fairness. 
Jain's fairness index \cite{astn_model:Jain} is used for the comparison.
Jain's Fairness Index is given by\\
\begin{equation}
\frac{ (\sum_{i=1}^{M} T_i )^2 }{ M \times \sum_{i=1}^{M} T_i^2 }
\end{equation}
where $ M $ is the number of STAs and $ T_i $ is the throughput of 
the $ i^{th} $ STA. The fairness index lies in the range 1 to $ 1 / M $.
When one STA obtains all the capacity, the fairness index equals $ 1/M $.
If the fairness index is 1, then all the STAs  achieve equal throughputs.
In Figure \ref{fig:fairness} we compare the fairness performance of the EDA policy 
with that of the others. It can be seen that  the EDA and Available Bandwidth based
policies achieve comparable degrees of fairness, which are distinctly better than those
 achieved by the others.

\section{Extensions}\label{sec:Extension}
We presented an association policy by considering 
TCP-controlled file downloads. This traffic constitutes a very substantial
part of network traffic,  as mentioned in \cite{astn_model:Businesswire}.
 Now let us consider simultaneous TCP uploads. In DCF, A node's attempt
  behaviour is independent of packet length.
If we interchange downlink data packets sent by APs with 
TCP-ACK packets and  uplink ACK packets sent by STAs with TCP data packets,
the same analysis holds good for TCP-controlled file uploads. The next case arises
 when some STAs are uploading and some STAs are downloading files. Here also our 
basic Markov model of the number of STAs with packets to send remains the same, 
if all the TCP windows are equal. Even different window sizes can be taken care of 
by finding the probability that the packet at the head of the queue at the AP is for
a particular STA. 

Uncontrolled UDP transfers bring changes in the model and
results. Apart from the above roaming is not addressed explicitly; it
can be considered as a departure from one AP and an arrival to another AP.
 We considered a noise free environment. Introducing link errors by 
considering noisy environments may change the actual throughput compared 
to the one which we estimate before association. In our simulation and numerical
evaluation, we used the 802.11b standard. However our mathematical
expressions are independent of this standard; hence the association policy can 
support $ k$ different data rates. Thus, we can apply the same scheme and analysis
for 802.11g/n standards. We assumed small RTTs, clearly, this needs to be generalized, and we are working on this at present. Further, we considered exponentially distributed files
with different mean sizes for different classes. As the BCMP network model 
supports file size distributions that are nonexponential also, our approach
can accommodate these; however, the details are to be worked out.

\section{Conclusion}\label{sec:Conclusion}
We developed an STA-AP association policy geared towards TCP-controlled file downloads. Our approach was based on extending a model for studying TCP-controlled infinitely long file transfers to the multirate WLAN context, and incorporating the output from this in a 2-node closed-queueing network model for estimating finite file download times. The resulting policy showed considerable improvement over other policies, ranging from 12.8\%
better throughput compared to EVA, up to 46.4\%
better throughput compared to simple RSSI-based association for a 9-AP network.

The association policy developed here is but an application of the models analyzed in earlier sections. The models and the analytical approach may find use in addressing other performance evaluation questions. For example, they could be used to address some ``what-if'' questions like: What would be the performance benefit that can be expected by a network upgrade that introduces a few new APs to of load the traffic from already-deployed congested APs?

% or


%\appendices

%\section{}

%\section*{Acknowledgement}


\ifCLASSOPTIONcaptionsoff
  \newpage
\fi



%\begin{biography}[{\includegraphics[width=1in,height=1.25in,clip,keepaspectr
% atio]{mshell}}]{Michael Shell}

%\begin{IEEEbiography}{}
%\end{IEEEbiography}

%\begin{IEEEbiographynophoto}


\end{document}